\documentclass[twocolumn,runinaddress,showpacs,prl]{revtex4}
\usepackage{amsmath}
\usepackage{graphicx,bm}

\begin{document}

\title{Transverse electric current induced by optically injected spin
current in cross-shaped InGaAs/InAlAs system}
\author{Jian Li, Xi Dai, Shun-Qing Shen, and Fu-Chun Zhang}
\affiliation{Department of Physics and Centre of Theoretical and Computational Physics,
The University of Hong Kong, Hong Kong, China}
\date{November 25, 2005}

\begin{abstract}
We examine electric response of a linearly polarized light normally shed on
a cross-shaped quasi 2-dimensional InGaAs/InAlAs system with structure
inversion asymmetry. The photo-excited conduction electrons carry a pure
spin current with in-plane spin polarization due to the Rashba spin-orbit
interaction. We use Landauer-B\"{u}ttiker formalism to show that this spin
current induces two inward or outward transverse charge currents, which are
observable in experiments. This effect may serve as an experimental probe of
certain types of spin current.
\end{abstract}

\pacs{72.25.-b, 75.47.-m}
\maketitle

Spin coherent transport of conduction electrons in semiconductor
heterostructures is currently an emerging subject due to its possible
application in a new generation of electronic devices \cite{Wolf01Science}.
There have been considerable efforts to achieve spin polarized current or
pure spin current in semiconductors such as injection from ferromagnetic
contact \cite{FM}, quantum spin pump \cite{Watson03}, spin Hall effect\cite%
{SHE}, and spin like Andreev reflection \cite{Wang05prl}. Optical injection
of spin current is largely based on the fact that the spin polarized
carriers in conduction band can be injected in semiconductors via absorption
of the circularly or linearly polarized light \cite{Ganichev, Bhat05prl}.
While there are successful ways to inject or generate spin current, its
detection is still a subtle problem. Despite the optical methods which
indicated spin accumulation due to the spin current \cite{SHEexperiments},
it is tempting to probe spin current by measuring its electric effects
\qquad \cite{Hirsch99prl, Zhang05arxiv, Hankiewicz05prb, Shen05prl}. In this
Letter, we theoretically study electric transverse current driven by an
optically injected spin current in a two-dimensional electron gas of
InGaAs/InAlAs with structure inversion asymmetry. A linearly polarized light
pumps electrons from valence to conduction bands, which induces a pure spin
current with in-plane polarization due to the spin orbit coupling. The Hall
effect related to this spin current in a cross-shaped mesoscopic system is
explored, which yields two measurable inward or outward electric transverse
currents. By using the Landauer-B\"{u}ttiker formalism we provide an
estimate of the electric transverse current measurable in experiments.

\begin{figure}[tbp]
\centering \includegraphics[width=0.4\textwidth]{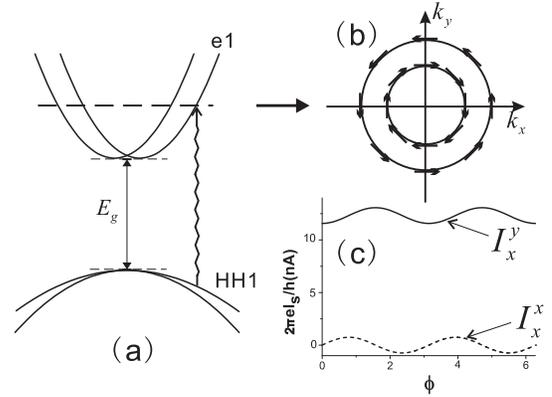}
\caption{(a) A schematic view of the 2D band structure of InGaAs/InAlAs. (b)
Scehmatic plot of spin polarization (arrow) of conduction electron in
momentum space with the same energy. (c) Calculated spin current induced by
linearly polarized light (see text for parameters).}
\label{fig:bands}
\end{figure}

We begin with the optical injection of spin current in quasi-2D
InGaAs/InAlAs with structural inversion asymmetry. The low energy band
structure is well known and plotted schematically in Fig.{\ref{fig:bands}}%
(a). The conduction electron can be described by an effective Hamiltonian,
\begin{equation}
H_{conduction}=\frac{\mathbf{p}^{2}}{2m_{c}}-\frac{\alpha }{\hbar }(\mathbf{%
p\times \sigma })\cdot \hat{z}+V(z)  \label{HE}
\end{equation}%
where $\sigma $ are the Pauli matrices, $V(z)$ is the asymmetric confining
potential perpendicular to the sample ($x$-$y$) plane, and $\alpha $ is the
strength of the Rashba spin-orbit coupling. The zero-field splitting of the
conduction band arises because of the Rashba coupling, and electrons are
spin polarized normal to the momentum in each subband, as shown in Fig.{\ref%
{fig:bands}}(b). The electrons in the valence band near the $\Gamma $ point
can be described by the Luttinger Hamiltonian,\ \
\begin{equation}
H_{valence}=(\gamma _{1}+\frac{5}{2}\gamma _{2})\frac{\mathbf{p}^{2}}{2m_{e}}%
+2\gamma _{2}\frac{(\mathbf{p}\cdot \mathbf{S})^{2}}{2m_{e}}+V(z)  \label{HH}
\end{equation}%
where $m_{e}$ is the free electron mass, $\gamma _{1}$ and $\gamma _{2}$ are
two Kohn-Luttinger parameters, which are taken to be $\gamma _{1}=7.0,$ and $%
\gamma _{2}=1.9$ in this calculation \cite{Dai05arxiv}, $\mathbf{S}$
represent three $4\times 4$ spin-$3/2$ matrices. For simplicity, we
approximate $V(z)$ in Eqs.(\ref{HE}) and (\ref{HH}) by an infinite potential
wall with a width of $d=10$nm. The wave functions for the holes in the
valence band can be obtained by a truncation method. We diagonalize $%
H_{valence}$ in a truncated Hilbert space only including the lowest N basis
states. In the present paper, we take $N=80$, which is accurate enough for
the lowest 4 hole subbands \cite{Dai05arxiv}.

The process of optical excitation is schematically illustrated with the e1
and HH1 subbands in Fig.{\ref{fig:bands}}. When a linearly polarized light
is shed normally onto the sample plane, the electrons are pumped from the
HH1 subband of the valence band to the e1 subband of the conduction band via
direct optical absorption provided that the photon energy is higher than the
band gap, i.e. $\hbar \omega >E_{g}$. Due to the Rashba spin-orbit coupling,
the conduction bands are split into two subbands as shown in Fig. 1(a) and
(b). A photo-excited electron has an in-plane spin polarization
perpendicular to its momentum, which induces a pure spin current. The spin
current operators for electrons in the conduction band and holes in the
valence band can be expressed in terms of the velocity and spin operators of
electrons and holes, respectively, $J_{\nu }^{\mu (e)}=\frac{\hbar }{4}%
\{\sigma ^{\mu },v_{\nu }^{(e)}\}$ and $J_{\nu }^{\mu (h)}=\frac{\hbar }{2}%
\{S^{\mu },v_{\nu }^{(h)}\}$, where $\mathbf{v}^{(e)}(\mathbf{k)}$ and $%
\mathbf{v}^{(h)}(\mathbf{k)}$ are the velocity operators for the conduction
and valence bands, respectively. The total spin current from the
photo-excited electrons and holes are,
\begin{equation}
\left\langle J_{\nu }^{\mu }\right\rangle =\sum_{\mathbf{k}}Tr\left\{ J_{\nu
}^{\mu (e)}\rho ^{(e)}(\mathbf{k})-J_{\nu }^{\mu (h)}\rho ^{(h)}(\mathbf{k}%
)\right\}
\end{equation}%
where $\rho ^{(e)}(\mathbf{k})$ and $\rho ^{(h)}(\mathbf{k})$ are
the density matrices for the conduction and valence bands,
respectively. The density matrices that appear in Eq. (3) can be
obtained in a relaxation time approximation. In the present study
only the diagonal components of the density matrices are kept and
the results can be written as \cite{transitionrate}
\begin{eqnarray}
\rho _{nn}^{(e,h)}(\mathbf{k}) &=&\frac{\pi }{2\hbar }\tau
_{e,h}\sum_{m}M_{nm}^{\ast }(\mathbf{k})M_{mn}(\mathbf{k})  \notag \\
&&\times \left[ \delta \left( E_{n}^{e,h}(\mathbf{k})+E_{m}^{h,e}(\mathbf{k}%
)-\hbar \omega _{ph}\right) \right]
\end{eqnarray}%
where $\tau _{e,h}$ is the relaxation time for electrons (holes) and $M$ is
the $2\times 4$ transition matrix describing the direct optical transition
between the conduction and valence subbands, which is caused by the external
light. Under the dipole approximation, $M$ can be expressed by $M=D_{c}^{-1}%
\widetilde{M}D_{v}$, where $D_{c}$ and $D_{v}$ are the $2\times 2$ and $%
4\times 4$ transformation matrices which diagonalize the Hamiltonian in Eqs.(%
\ref{HE}) and (\ref{HH}), respectively, and the matrix $\widetilde{M}$ is
the coupling matrix in the original basis of the row $\left\{ \left\vert
S_{z}\left( =+1/2,-1/2\right) \right\rangle \right\} $ and column $\left\{
\left\vert S_{z}\left( =+3/2,1/2,-1/2,-3/2\right) \right\rangle \right\} $ as%
\begin{equation}
\widetilde{M}=\left(
\begin{array}{cccc}
ge^{i\phi } & 0 & g^{\ast }e^{-i\phi }/\sqrt{3} & 0 \\
0 & ge^{i\phi }/\sqrt{3} & 0 & g^{\ast }e^{-i\phi }%
\end{array}%
\right)
\end{equation}%
where $\phi $ is the polarization angle of the linearly polarized light and
the factor $g$ is determined by the Bloch functions of electron and hole at
the $\Gamma $ point. In this way the pure spin current is obtained as a
function of the frequency and polarization of the light. The dominant
component of the spin current flowing in the $x$ direction is $J_{x}^{y}$,
which is similar to the equilibrium current proposed by Rashba \cite%
{Rashba03prb}. To calculate the spin current we adopt the parameters from a
sample of In$_{x}$Ga$_{1-x}$As/In$_{0.52}$Al$_{0.48}$As \cite{Yang05}, with
the Rashba coupling strength $\alpha =6.1\times 10^{-12}$eVm, ($\alpha
/\hbar \simeq 3\times 10^{-4}c$, where $c$ is the speed of light,) the
effective mass $m_{c}=0.05m_{e}$, the incident light power is $100$mW with
the wavelength $\lambda =880$nm. Also we extract from the experimental data
\cite{Yang05} that ${\frac{{\pi |g|^{2}\tau _{e}}}{{(\hbar ^{2}/{m_{e}d^{2}}%
)2\hbar }}}\approx 0.78\times 10^{-3}$, which is used in calculating the
density matrices. Given a quantum well with the size to be $L\times L$, the
induced current $I_{\nu }^{\mu }=J_{\nu }^{\mu }L$ varies approximately with
$\cos 2\phi $, as shown in Fig.\ref{fig:bands}(c) with $L=100\mu m$, the
function of $I_{x}^{y}$ is fitted to be $I_{x}^{y}\approx I_{0}+I_{1}\cos
2\phi $ with $I_{0}=12.32nA$ and $I_{1}=0.75nA$. And it is noticeable that
there is also a non-vanishing $I_{x}^{x}$ component of the spin current,
although it's comparatively negligible.

We now turn to investigate the consequence of applying this in-plane
polarized spin current to a cross-shaped mesoscopic system with the Rashba
spin-orbit coupling, and it turns out that electric transverse currents will
be induced in this case. Inevitably this is a situation reminiscent of the
reciprocal version of the spin Hall effect proposed by Hirsch \cite%
{Hirsch99prl} and Zhang et al \cite{Zhang05arxiv}, in which a transverse
current was predicted to arise when a spin current polarized \emph{%
perpendicular} to the plane flows through the scattering region. And a
general Onsager relation between the spin Hall conductance and the electric
Hall conductance has been found in a similar condition \cite{Hankiewicz05prb}%
. While the spin current is indeed a tensor, not a conventional vector, its
spin polarization as well as the symmetries of the scattering spin-orbit
coupling play key roles in generating the electric transverse current, which
implies that for \emph{in-plane} polarized spin current, the induced
transverse currents may have a quite different character. Here we show this
difference from the symmetry point of view, estimate the amplitude of
transverse current numerically, and then propose an experiment to observe
the transverse electric currents induced by the in-plane polarized spin
current in our case.

\begin{figure}[tbp]
\centering \includegraphics[width=0.4\textwidth]{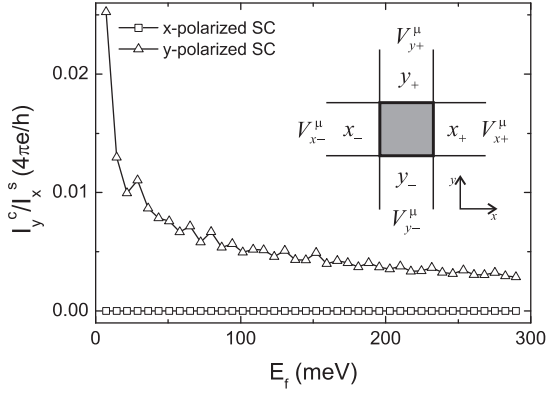}
\caption{The ratio of the electric transverse current to the longitudinal
spin current as a function of the Fermi energy counted from the bottom of
the conduction band, with inset the geometry of the cross-shaped structure
in our calculation. The calculation was carried out on a $40\times 40$
square lattice with the total width $L=100$nm, the effective electron mass $%
m^{\ast }=0.05m_{el}$, and the Rashba spin-orbit coupling strength $\protect%
\alpha =6.1\times 10^{-12}$eVm.}
\label{fig:cross}
\end{figure}

In our proposal the whole setup is carved on an InGaAs/InAlAs heterojunction
with a central scattering region and four leads for measurements, as shown
in the inset of Fig. \ref{fig:cross}. The transverse leads (leads $y_{+}$
and $y_{-}$) and the scattering region should be masked to prevent from the
light shining explicitly, while the longitudinal leads (leads $x_{+}$ and $%
x_{-}$) are opened to accept the linearly polarized light to generate the
incident in-plane polarized spin current. By avoiding any possible
interfaces of current injection, the spin current is expected to circulate
in the $x$ direction without conventional spin injection problem. Under the
Landauer-B\"{u}ttiker formalism \cite{Buttiker86PRL, Datta95}, which has
been extensively used in the study of quantum transport in mesoscopic
systems \cite{LBspinHall, LBsymmetry}, and will be applied to analyze the
transverse currents in our case, we simulate the generation of the spin
current and the measurement of the transverse currents by assuming a proper
setting of the spin-dependent chemical potential related to each lead. In
detail, if we denote the effective voltage related to spin polarization $\mu
$ ($\mu =\uparrow ,\downarrow $) at lead $p$ by $V_{p}^{\mu }$, then we
assume $V_{x-}^{\uparrow }=-V_{x-}^{\downarrow }=-V_{x+}^{\uparrow
}=V_{x+}^{\downarrow }=V_{0}/2$ and $V_{y-}^{\uparrow }=V_{y-}^{\downarrow
}=V_{y+}^{\uparrow }=V_{y+}^{\downarrow }=0$. It should noted that $\mu $
can be oriented in $x$, $y$ or $z$ direction (to consider either in-plane or
perpendicular-to-plane polarized spin current), which will be denoted by $%
\mu \sim x,y$ or $z$. Given this voltage setting, the symmetry properties of
the currents will be fully presented by the spin-dependent transmission
functions between the leads \cite{LBsymmetry}, as we will see.

In each lead (assumed to be ideal and semi-infinite by convention) attached
to the central shaded region in the inset of Fig. 2, the wave function can
be expanded in terms of separate propagating modes, which are the
eigenstates in the lead. In $x_{-}$, for example, the eigenstates are $\psi
_{m\mu }^{\pm }(x,y)=C\,e^{\pm ik_{m}x}\phi _{m}(y)\otimes \chi _{\mu }$,
where $C$ is the normalization constant, $\phi _{m}(y)$ is the $m^{th}$
eigenstate in the transverse dimension, $\chi _{\mu }$ is the spin
eigenstate with $\mu =\uparrow $ or $\downarrow $ and $\pm $ denotes the
mode is incoming or outgoing. The expansion coefficients of the actual wave
functions in terms of these eigenstates are known as the wave amplitudes $%
a_{m\mu }^{in/out}$, which are related by an unitary matrix $a_{m\mu
}^{out}=\sum_{n,\nu }s_{mn}^{\mu \nu }a_{n\nu }^{in}$ and all the symmetries
of the transport properties in the system are embedded in this S-matrix. For
simplicity we assume that the quantum modes in opposite leads are
symmetrical, which means that for each mode with wave vector $k_{m}$ and
spin polarization $\mu $ in lead $x_{-}$($y_{-}$) there is a mode with the
same wave vector and spin polarization in lead $x_{+}$($y_{+}$), and vice
versa. In the clean limit, the Hamiltonian with the Rashba spin-orbit
coupling term is obviously invariant under three unitary transformations,
i.e. $H$ commutes with the time-reversal operator $T=-i\sigma _{y}K$, where $%
K$ is the complex-conjugate operator, and two combined operators $\sigma
_{x}R_{x}$ and $\sigma _{y}R_{y}$, where $R_{x}$($R_{y}$) denotes the mirror
reflection operator transforming $x\rightarrow -x$ ($y\rightarrow -y$).
While the transformed eigenstates remain as eigenstates of the original
Hamiltonian, some of the amplitudes are transformed in the following ways:%
\newline
under the transformation of $T$,
\begin{equation}
a_{m\mu }^{in}\rightarrow (a_{m\bar{\mu}}^{out})^{\ast };\;a_{m\mu
}^{out}\rightarrow (a_{m\bar{\mu}}^{in})^{\ast };  \label{amp_t}
\end{equation}%
under the transformation of $\sigma _{i}R_{i}$,
\begin{eqnarray}
a_{m\mu }^{in(out)}\rightarrow a_{\bar{m}\mu }^{in(out)} &&(m\in
i_{-},i_{+};\mu \sim i)  \notag \\
a_{m\mu }^{in(out)}\rightarrow a_{\bar{m}\bar{\mu}}^{in(out)} &&(m\in
i_{-},i_{+};\mu \sim j\neq i);  \label{amp_x}
\end{eqnarray}%
where $i=x$ or $y$, $\bar{\mu}=-\mu $ and $\bar{m}$ is the counterpart of $m$
in its opposite lead. The phase factors are neglected in the above
transformations because they will not be manifested in the following
calculations of the transmission probabilities. The symmetries as well as
the unitary condition impose constraints on the S-matrix, and thus on the
transmission probability from mode $\{n,\nu \}$ to mode $\{m,\mu \}$, which
is defined as $T_{mn}^{\mu \nu }\equiv |s_{mn}^{\mu \nu }|^{2}$. For the
time reversal symmetry $T$, this yields
\begin{equation}
T_{mn}^{\mu \nu }=T_{nm}^{\bar{\nu}\bar{\mu}};  \label{tmat_t}
\end{equation}%
for the symmetry under $\sigma _{i}P_{i}$,
\begin{eqnarray}
T_{mn}^{\mu \nu } &=&T_{m^{\prime }n^{\prime }}^{\mu \nu }\quad (\mu ,\nu
\sim i)  \notag \\
T_{mn}^{\mu \nu } &=&T_{m^{\prime }n^{\prime }}^{\bar{\mu}\bar{\nu}}\quad
(\mu ,\nu \sim j\neq i)  \label{tmat_x}
\end{eqnarray}%
where $i=x$ or $y,\;m^{\prime }=\bar{m}$ if $m\in i_{-},i_{+}$, or $m$
otherwise. By summing up all the transmission probabilities between two
leads with specific spin polarizations, the transmission functions $\bar{T}%
_{pq}^{\mu \nu }=\sum_{m\in p,n\in q}T_{mn}^{\mu \nu }$, and the currents
are obtained using the extended B\"{u}ttiker formula \cite{Buttiker86PRL} $%
I_{p}^{\mu }=\frac{e}{h}\sum\limits_{q,\nu }(\bar{T}_{pq}^{\mu \nu
}V_{q}^{\nu }-\bar{T}_{qp}^{\nu \mu }V_{p}^{\mu })$ with the electric and
spin current defined as $I_{p}^{c}=e(I_{p}^{\uparrow }+I_{p}^{\downarrow })$
and $I_{p}^{s}=\frac{\hbar }{2}(I_{p}^{\uparrow }-I_{p}^{\downarrow })$,
respectively. Combining the symmetry-derived Eqs.(\ref{tmat_t}) and (\ref%
{tmat_x}) with the preceding voltage configuration, we find
\begin{subequations}
\begin{eqnarray}
&&%
\begin{array}{cc}
I_{y-}^{c}=-I_{y+}^{c}=0, & (\mu \sim x)%
\end{array}%
\quad   \label{c4x} \\
&&\quad
\begin{array}{cc}
I_{y-}^{c}=I_{y+}^{c}, & (\mu \sim y)%
\end{array}
\label{c4y} \\
&&\quad
\begin{array}{cc}
I_{y-}^{c}=-I_{y+}^{c}, & (\mu \sim z)%
\end{array}
\label{c4z}
\end{eqnarray}%
It is clear now to see the difference between the transverse electric
currents induced by a $z$-direction polarized spin current and an in-plane
polarized spin current, that is in the former case the transverse current is
a truly circulating one, which can be naturally regarded as a reversed
effect of the spin Hall effect \cite{Hirsch99prl, Zhang05arxiv,
Hankiewicz05prb}, whereas in our case of in-plane polarization, the
transverse currents are flowing both inward or outward instead of
circulating through the two transverse leads. Consequently this will make an
essential difference in the measurements of the currents in such systems. To
present a quantitative estimate of the induced currents, we make numerical
calculations with the tight-binding approximation in this cross-shaped
mesoscopic system, and the ratio of the induced electric transverse current $%
I_{y}^{c}$ to the spin current $I_{x}^{s}$ are plotted in Fig.\ref{fig:cross}%
. Combined with the calculated value of the injected spin current, we notice
that the induced current is about $0.1\sim 0.2$nA, which is large enough to
be measured experimentally, while the small part of the spin current with
spin polarization along $x$ axis does not contribute to the transverse
currents.

\begin{figure}[tbp]
\centering \includegraphics[width=0.45\textwidth]{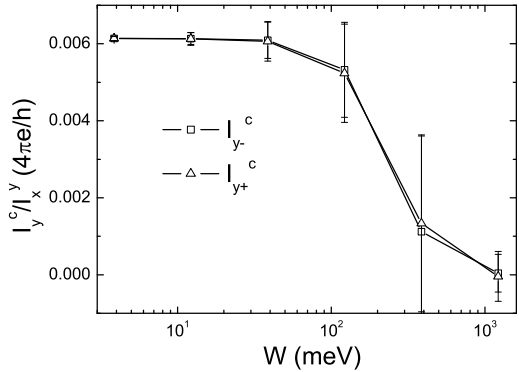}
\caption{The ratio of the electric transverse current to the longitudinal
spin current polarized in y-direction as a function of the disorder strength
$W$, with the error-bars showing the standard deviations. The calculation
was carried out on a 30$\times $30 square lattice with the total width $L=75$%
nm, the Fermi energy $E_f=62$meV (counted from the bottom of the conduction
band), the effective electron mass $m^{\ast }=0.05m_{el}$, and the Rashba
spin-orbit coupling strength $\protect\alpha =6.1\times 10^{-12}$eVm.}
\end{figure}

In practice, there are two major factors which will affect our conclusions,
the disorder effect and the absence of the symmetry of leads we've assumed
previously. The disorder breaks the symmetries of the system in a way as
adding some random on-site potentials. Our numerical calculation with
on-site potentials varying within $[-W/2,W/2]$, illustrated in Fig. 3, shows
that despite the current approaching zero with increasing $W$, the symmetry
relations shown in Eq.(\ref{c4x}) and Eq.(\ref{c4y}) are well preserved in
terms of the average values provided $W$ is not too large. This is
reasonable and consistent with previous works \cite{LBspinHall}. As for the
absence of the symmetry of the quantum channels in the leads, we mimic its
effect by modifying the width of each lead, and the calculations show that
the profile of each current is retained with its value changed according to
the modification of the channel number, which implies that within a limited
range of asymmetry, our conclusions are still applicable.

In conclusion, a linearly polarized light may pump electrons from the
valence band to the conduction band in InGaAs/InAlAs heterojunction, and the
zero-field splitting of the conduction band with structural inversion
asymmetry drives the photon-excited electrons and holes to form the spin
current with in-plain spin polarization. The spin current was estimated by
means of the electric-dipole approximation and relaxation time
approximation. This mechanism provides a source of spin current to explore
spin-dependent transport in mesoscopic systems. Furthermore, the Landauer-B%
\"{u}ttiker formula is used to calculate the electric Hall currents while
optical injection of spin current is empirically regarded as a source of
spin-dependent potentials. As a result two inward or out electric transverse
currents are observed. All parameters in our calculation are adopted from a
realistic sample of InGaAs/InAlAs heterojunction. Thus the effect may serve
as an experimental probe of in-plane polarized spin currents.

The authors thank J. Sinova for helpful discussions. This work was supported
in part by the RGC of Hong Kong under Grant No. HKU 7039/05P.

\end{subequations}

\end{document}